\documentstyle[editedbook,epsfig,psfig,epsf]{mq}

\def\cm2{cm$^2$ }
\def\se1{s$^{-1}$ }

\begin{opening}
\title{GRS 1915+105: Ten Years After}
\author{T. Belloni}
\institute{INAF--Osservatorio Astronomico di Brera, Via E. Bianchi 46,
           I-23807 Merate (LC), Italy.}
\end{opening}

\runningtitle{GRS 1915+105: Ten Years After}
\runningauthor{Belloni}

\begin{document}
\vspace{-0.5cm}
\begin{abstract}
{\small 
It is now ten years since the first microquasar GRS 1915+105 was discovered.
More than six years of observations with RossiXTE have shown a level 
of variability never observed in any other X-ray source. Here I try to 
address some issues, based on X-ray observations only,
that have relevance for theoretical modeling.
First, I ignore these peculiarities and concentrate on the similarities
with other X-ray transients. Then I focus on the peculiar variability and
present a number of obervational facts that need to be addressed by
theoretical models.
}
\end{abstract}

\section{GRS 1915+105 from GRANAT to RossiXTE}

As practically all articles about GRS 1915+105 start, the source was discovered
ten years ago, in 1992,
with the Watch instrument on board GRANAT \cite{castro}. Figure 1 
shows the original discovery light curve from that work, where one can see
that the source was very variable from the beginning.
As a comparison, the
full RossiXTE/ASM light curve up to 2002 July is shown in the bottom panel
of Fig. 1. Already from these plots, it is evident that the variability of
GRS 1915+105 is rather unique and that in this source we are observing something
which is not observed in other sources.

\begin{figure}[htb]
\centering
\psfig{file=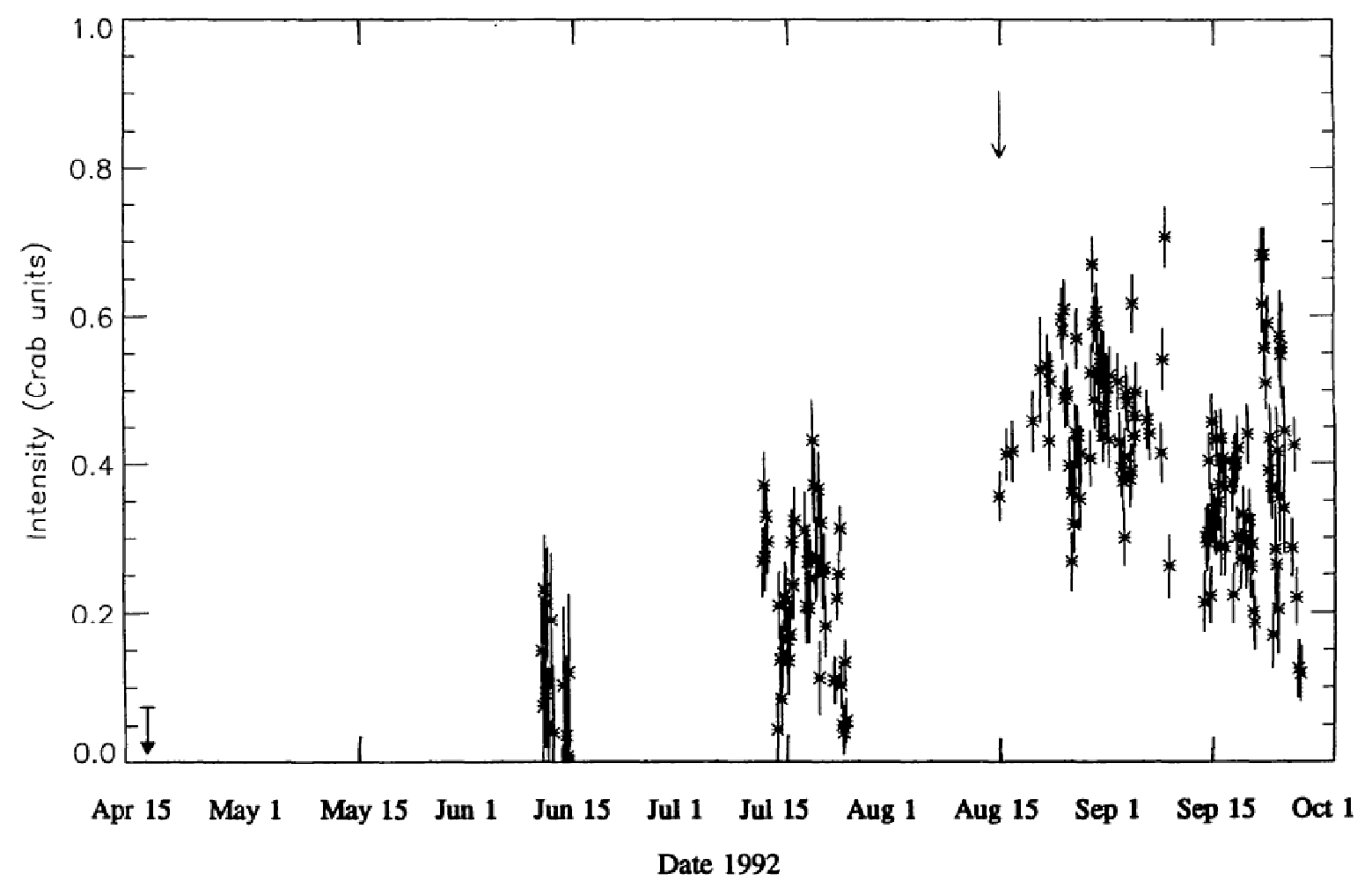,width=10cm}
\psfig{file=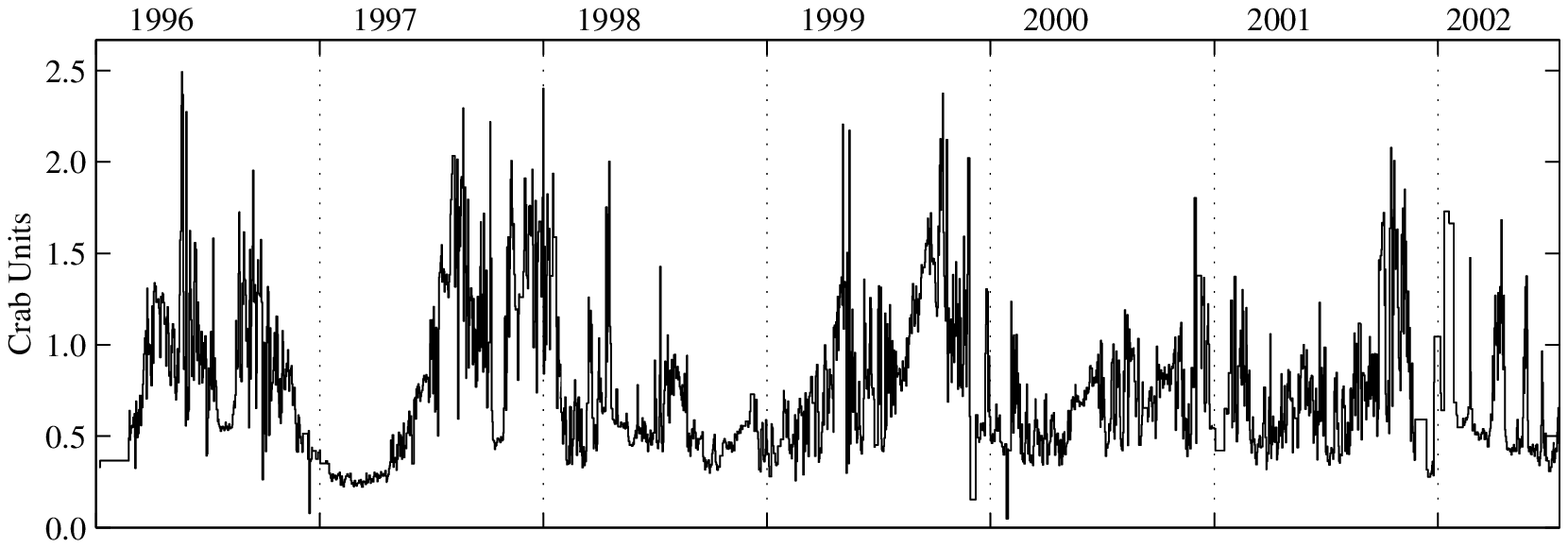,width=10cm}
\caption{GRS 1915+105 long-term light curves. Top: GRANAT/Watch 
         [1]. Bottom: RossiXTE/ASM up to 2002 July 19.}
\label{fig:ex}
\end{figure}

When examining the variability at shorter time scales, as observed with the
RossiXTE/PCA, things become even
more complex (see Fig. 2), showing a definite structure, which in this
particular case repeats almost unchanged after about 25 minutes.  
The complexity of these light curves, first shown by
\cite{gmr97}, is analyzed and categorized by \cite{b2000}, who classify the
light curves in a dozen variability classes and identify three basic
spectral states called A, B and C.
In these ten years, 
many papers have been devoted to the analysis and interpretation of the
X-ray observations of GRS~1915+105, and it is not possible to review them
all here. What I want to do is to put this unique source in perspective,
comparing it with other transient BHCs. First, I will emphasize the 
similarities with other systems, then I will concentrate on the
peculiarities.

\begin{figure}[htb]
\centering
\psfig{file=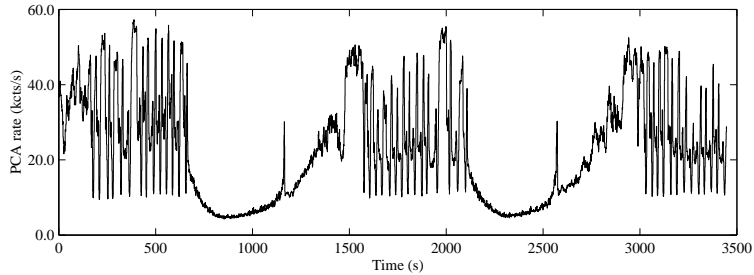,width=10cm}
\caption{Example of a RossiXTE/PCA light curve of GRS 1915+105, from 1997
	October 31.}
\label{fig:ex}
\end{figure}

\section{GRS 1915+105 as a transient}

\begin{figure}[htb]
\centering
\psfig{file=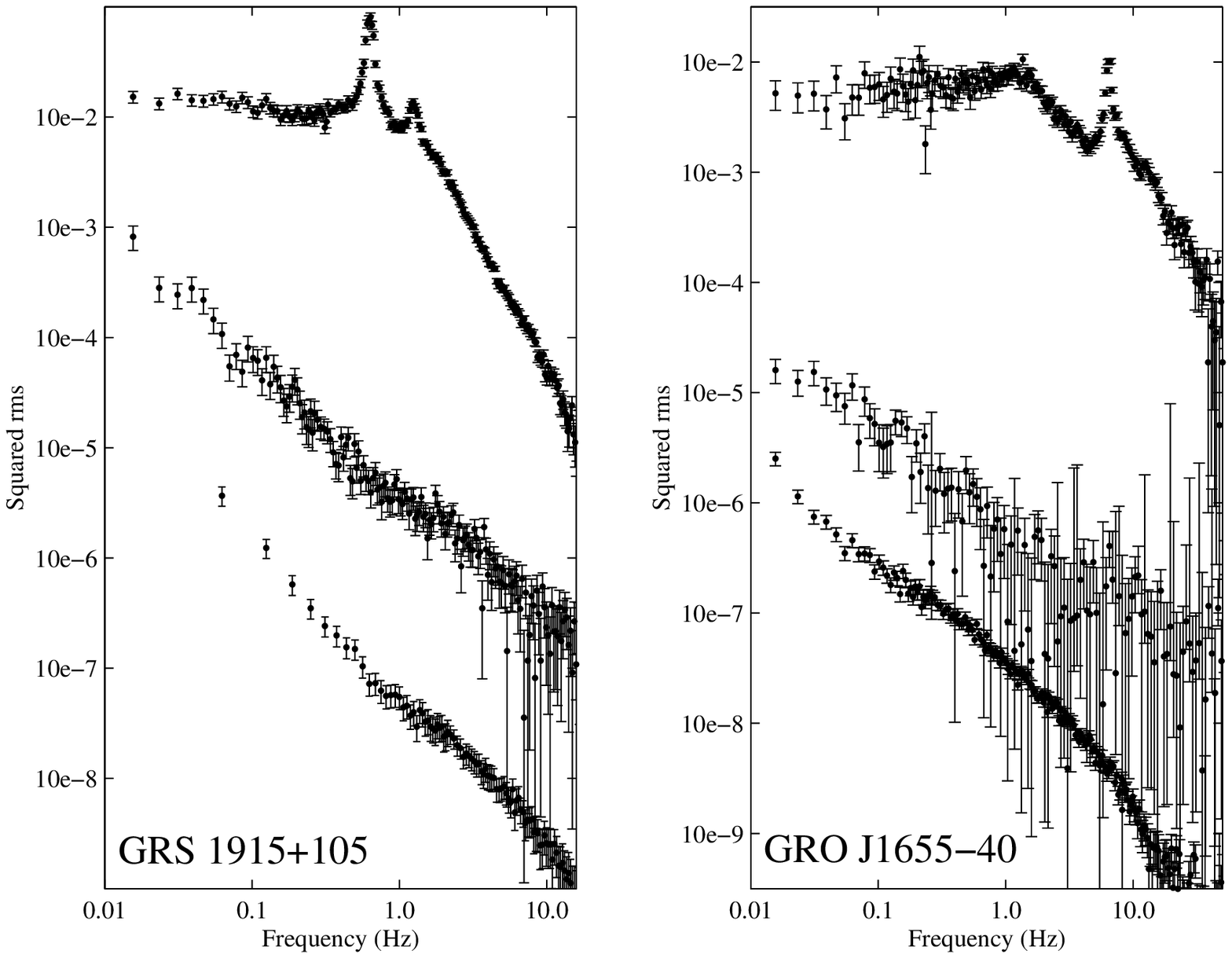,width=10cm}
\caption{Left panel: PDS of the three basic spectral states
	of GRS 1915+105 (C, A and B from top to bottom, shifted by a factor
	of 100 from each other). Right panel: PDS of three
	observations from the latest stages of the outburst of GRO~J1655-40,
	shifted in the same way.}
\label{fig:ex}
\end{figure}

Looking at GRS~1915+105 as an X-ray transient, if we ignore the most
obvious peculiarities, we can identify some common features which the system
shares with all other BH systems. One is the shape of the Power
Density Spectra (PDS) above, say, 1 Hertz (see e.g. \cite{mgr97,reigold}).
A second one is the shape of the broad-band energy spectra as
observed with BeppoSAX and RossiXTE: the presence of a soft thermal 
component plus a hard component that at times can extend to very high
energies (see e.g. \cite{bel97a,bel97b,muno,zdz}).

In order to compare the shape of the PDS, let us consider three representative
cases, corresponding to time intervals when the source is not wildly 
variable as in Fig. 2, corresponding to the A, B and C states identified
by \cite{b2000} (Fig 3, left panel). They can be compared to the PDS observed
from the microquasar GRO~J1655-40, which are typical from
the so-called canonical states of BHC \cite{men97}. One can see that the
similarities are rather strong, indicating a connection between the states
of GRS~1915+105 and the ``canonical'' states of other BHCs. 
A more detailed analysis is presented in \cite{reig02}.

Turning to the energy spectra, a comparison based on X-ray colors is also
considered in
\cite{reig02}. However, more interesting here is to notice that the 1-10 Hz
QPO observed in the C state of GRS~1915+105 follows the same correlation
with the slope of the power-law component in the energy spectrum which is
observed in a number of other sources \cite{vig02}. The correlation can 
be see in Fig. 4 for GRS~1915+105 and for the bright transient 
XTE~J1550-564. Also in this case there are striking similarities, including
the indication of a turnoff of the relation at high QPO frequencies.

\begin{figure}[htb]
\centering
\psfig{file=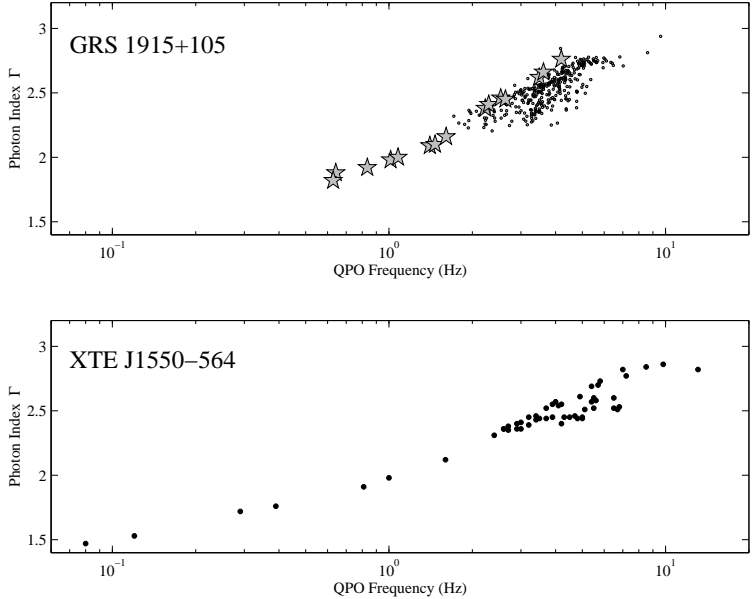,width=10cm}
\caption{Top panel: QPO-$\Gamma$ correlation for selected C state intervals
	of GRS~1915+105 (stars indicate $\chi$-class observations).
	Bottom panel: the same correlation observed in XTE~J1550-564.}
\label{fig:ex}
\end{figure}

All this indicates that, when ignoring intervals of strong and peculiar
variability, GRS~1915+105 behaves in a rather similar way to other
transient systems and, if these were the only available data, would be 
indistinguishable from them.

\section{GRS 1915+105 as a peculiar source}

Of course it is not really possible to ignore the most interesting parts of the
observations of this source. Therefore, let's turn to the peculiarities of
GRS~1915+105 and consider what we can learn from it. 
The observed oscillations in flux and energy spectrum, occurring on time
scales from a few minutes to months, have been interpreted as due to the
onset of an instability which causes the innermost region of the accretion
disk to switch between a visible and a non-visible state \cite{bel97a,bel97b}.
This idea has been recently developed theoretically by a number of authors, who
tried to reproduce the observed light curves with some success
\cite{naya,ewa,janiuk,janiuk2}. However, only the most basic features of the
observations are taken into account by these works.

A more detailed observational approach has been followed by \cite{muno}, who
analyzed a large number of spectra of GRS~1915+105 and presented correlation
between timing and spectral parameters. 
Recently, a complete analysis of a large number of
RossiXTE/PCA observations, with the production of energy spectra on the time
scale of 16 seconds, has been presented by \cite{migliari}. An example of this
analysis is shown in Fig. 5, where an observation interval is shown, together 
with the time history of the spectral parameters. 
A clear evolution in spectral parameters is observed.
These parameters can be
correlated with each other and can give important clues to detailed
theoretical models as those mentioned above. 
An important example is shown in Fig. 6. Here the best-fit 
inner radius and temperature of the accretion disk are plotted versus each
other (notice that although the absolute value of the inner disk cannot
be taken at face value, its variations are a more solid measurement
\cite{merloni}).The points, corresponding to different $\chi$ (state C only)
observations, follow a power-law correlation with index $\alpha\sim -$0.4,
much flatter than the value $\alpha = -3/4$ expected from a Shakura \&
Sunyaev disk at constant mass accretion rate. This indicates that moving from 
large radii to smaller radii, the measured local accretion rate 
through the disk at that
radius decreases. The possible association of this decrease in local
accretion rate and the generation of jets should be considered.

\begin{figure}[htb]
\centering
\psfig{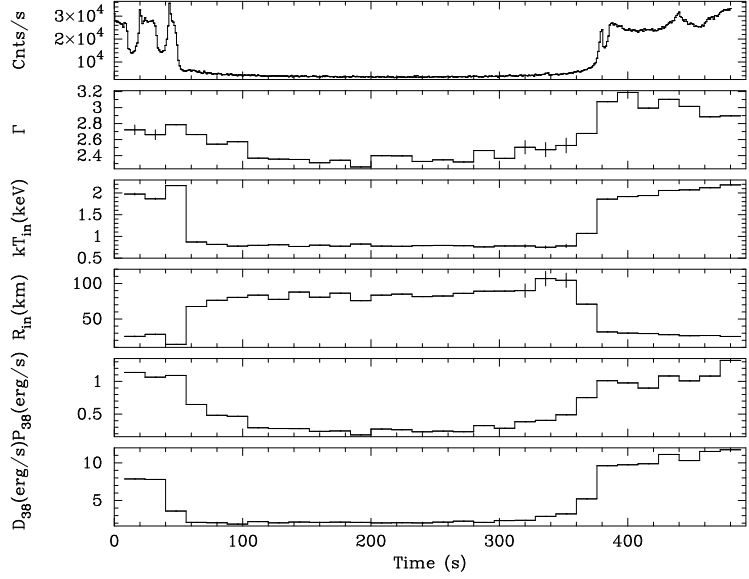}
\caption{Time evolution of spectral parameters of an observation interval
	of GRS~1915+105 as observed by the RossiXTE/PCA. Panels are (top to
	bottom): PCA light curve, power-law photon index $\Gamma$,
	disk inner temperature and radius, flux of the power-law and the
	of the disk component.}
\label{fig:ex}
\end{figure}

\begin{figure}[htb]
\centering
\psfig{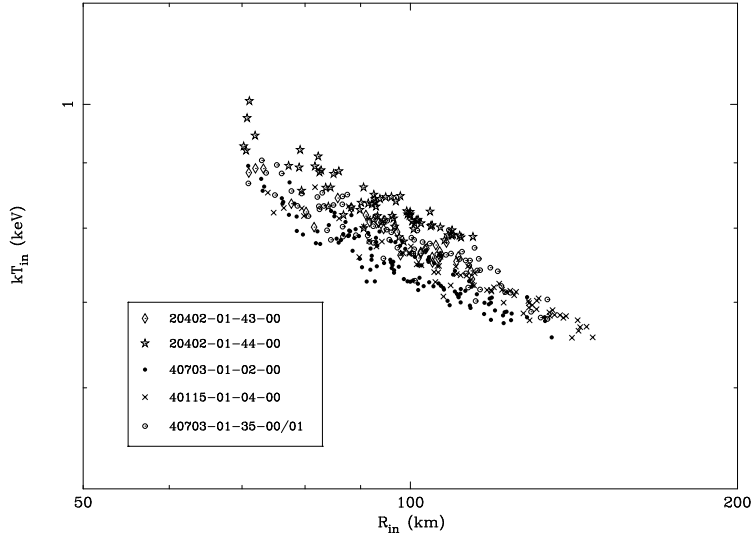}
\caption{Correlation between inner disk temperature and radius for a
	selected subsample of C-state intervals of GRS~1915+105 observed
	by RossiXTE/PCA. Time resolution is 16 seconds.}
\label{fig:ex}
\end{figure}

To put these variations in a broader perspective, we can expand the plot
of Fig. 6 and include the corresponding state-A and state-B intervals.
This is done in Fig. 7, where also two lines at constant mass accretion
rate are plotted. In addition to the $\alpha\sim -$0.4 described before
for state C (notice that this branch is followed twice, in opposite 
directions), we can see that as the radius decreases further (state A),
the points follow precisely a $\alpha=-3/4$ correlation, indicating that here
the local mass accretion rate is indeed constant. Notice that the soft state A
takes place {\it after} the radio event starts, indicating that state A comes
after a jet emission \cite{klein}. After state A, the source enters state B,
where the radius has reached its minimum value and the temperature slowly
increases, thereby increasing again the local mass accretion rate (see
also \cite{bel97a,b2000}). Once again, the behavior shown in Fig. 7 
needs to be addressed together with the shape of the light curves by any
theoretical model that aims at the interpretation of the peculiarities of
GRS~1915+105.

\section{The biggest challenges}

As I have shown in the previous sections, there are a number of detailed
observational results that can provide inputs to theoretical models for
the understanding of GRS~1915+105, and for connecting them to what is
observed from other BHCs. 
In this last section I would like to show two of the most outstanding
features of this source, features which I consider as the biggest challenges for 
any theoretical model to be applied to the peculiarities of this source. 
The first is the presence of extreme variability on time scales down to
0.2 seconds (see Fig. 8, top panel). This variability is accompanied by changes
in the X-ray colors, i.e. the spectrum softens when the flux decreases
\cite{b2000}. These changes can be attributed to changes in the temperature
of the inner disk, and therefore to variations in the local mass accretion 
rate. This can happen on a local viscous time scale, which is in the 
observed range. The cause  of this variability, spectrally different from 
the disk oscillations commonly considered, needs to be understood.

\begin{figure}[htb]
\centering
\psfig{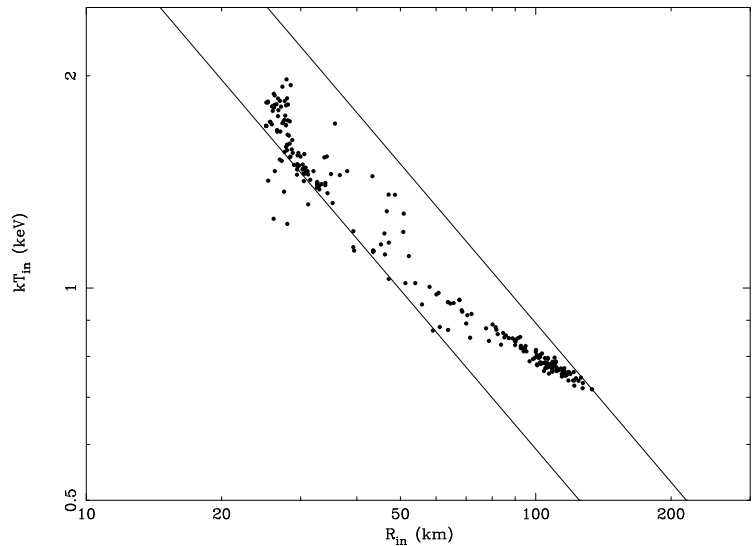}
\caption{Same as Fig. 6, but including the corresponding A-state and B-state
	intervals. The two diagonal lines correspond to lines of constant
	mass accretion rate for a Shakura \& Sunyaev disk.}
\label{fig:ex}
\end{figure}

The second challenge is the extreme degree of similarity between light curves
(and corresponding spectral evolution) of observations from different
variability classes (see \cite{b2000}). 
The dozen pattern that have been identified
in the light curves of GRS~1915+105 might seem many, but in reality they 
are surprisingly few, considering all possibilities for the oscillations. 
The fact that the source locks itself in this relatively small set of
patterns might be a key for the understanding of the origin of the
variability. Moreover, not only the general shape of the light curve,
but also specific details can be found in observations years apart. 
As an example (from \cite{b2001}), in the bottom panels of Fig. 8 I show
two light curves well distant in time where the pattern of variability
is followed even to some (albeit not all) details. Notice that one of
the two light curves (the top one) has been rescaled in time in order to
match the time scale of the other, indicating that some time scale 
underlying the whole process must have been different. 
Of course it is not realistic to ask for a theoretical model that 
explains and reproduces the exact shape of the light curves in Fig. 8.
However, the simple fact that such complex shapes are in some way
characteristic of the system should be taken into account when developing
models. For instance, models like that of \cite{janiuk2}, which produce
encouraging light curves, should as a next step keep into account
more details of the observations (like those in Fig. 8) without running
the risk of trying to interpret noise in the system.

\begin{figure}[htb]
\centering
\psfig{file=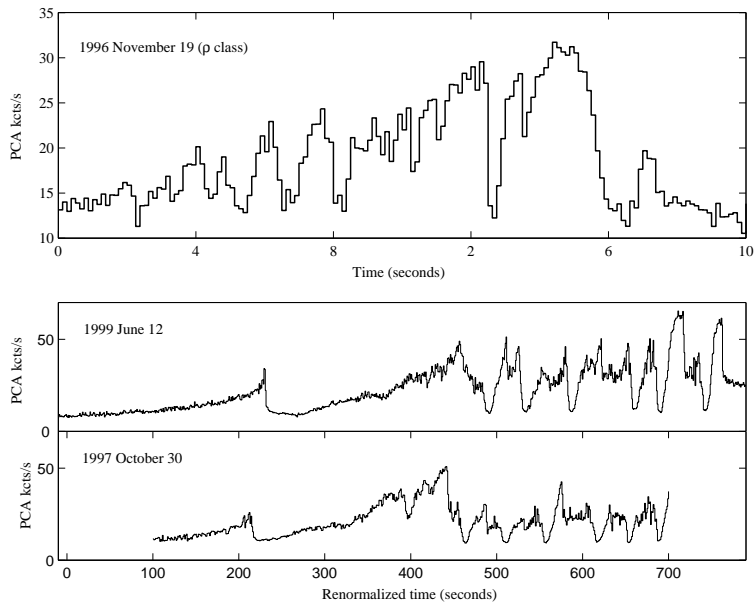,width=10cm}
\caption{Top panel: PCA light curve from a $\rho$-class observation 
       with 1/8 s binning. Very large-amplitude and fast variability is evident.
	Bottom panels: two PCA light curves (1s binning) from almost two
	years apart. The topmost one has been rescaled in time to highlight
	similarities in the shape.}
\label{fig:ex}
\end{figure}

\section{Conclusions}
   
GRS~1915+105 is a peculiar black-hole transient in at least two ways.
First, it can hardly be considered a transient, since it is in 
``outburst'' since ten years at large values of the accretion rate.
Second, as shown before, the variability of its flux and spectral
properties are unique and extremely complex. However, not all its properties
are unique (as in the case of XTE J0421+56, see \cite{bel99,par2000,boi02}),
indicating that we are not dealing with a source completely different from 
the others. This can be turned to our advantage: by linking ``normal'' and
peculiar characteristics, we can learn something general about accretion
onto black holes.

\section*{Acknowledgments}
I thank the Cariplo Foundation for financial support.

\end{document}